\documentclass[%
 reprint,
 amsmath,amssymb,
 aps,
]{revtex4-1}
\usepackage{bbm}

\usepackage{graphicx}
\usepackage{dcolumn}
\usepackage{bm}
\usepackage{amssymb}
\usepackage{amsmath,amssymb,mathrsfs,dsfont}
\DeclareMathOperator*{\argmax}{arg\,max}

\makeatletter

\newcommand{\Rmnum}[1]{\expandafter\@slowromancap\romannumeral #1@}
\makeatother

\begin{document}

\preprint{APS/123-QED}

\title{Dynamical Immunization Strategy for Seasonal Epidemics}

\author  {Shu Yan $^{1}$}
\email{yanshu@smss.buaa.edu.cn}
\author {Shaoting Tang$^{1}$}
 \email{tangshaoting@buaa.edu.cn}
\author{Sen Pei $^{1}$}
\author{Shijin Jiang$^{2}$}
\author{Zhiming Zheng$^{1}$}%
 \email{zzheng@pku.edu.cn}
\affiliation{%
 $^{1}$LMIB and School of Mathematics and Systems Science, Beihang University, Beijing 100191, China
\\$^{2}$School of Mathematical Sciences, Peking University, Beijing, 100871, China}%

\date{\today}

\begin{abstract}
The topic of finding effective strategy to halt virus in complex network is of current interest. We propose an immunization strategy for seasonal epidemics that occur periodically. Based on the local information of the infection status from the previous epidemic season, the selection of vaccinated nodes is optimized gradually. The evolution of vaccinated nodes during iterations demonstrates that the immunization tends to locate in both global hubs and local hubs. We analyze the epidemic prevalence by a heterogeneous mean-field method and present numerical simulations of our model. This immunization performs superiorly to some other previously known strategies. Our work points out a new direction in immunization of seasonal epidemics.

\end{abstract}

\pacs{Valid PACS appear here}
\maketitle


\section{\label{sec1}Introduction}
Epidemic spreading in complex networks has received considerable attention for the last two decades \cite{WS,Spreading,Newman2002, Newman2003,YS,Barrat2008, LWH,PS}. The immunization strategy to halt virus is an important topic in epidemic spreading research, and most previous literature of this field have focused on the selection of vaccinated nodes before the outbreak of an epidemic \cite{Imm1,Imm2,Imm3,Imm4,Imm5,Imm6,Imm7,Imm10,Imm11,Imm12,Imm13}. Numerous immunization strategies have been proposed such as uniform immunization (nodes are vaccinated randomly) \cite{Imm1,Imm6} or targeted immunization (most highly connected nodes are vaccinated) \cite{Imm2,Imm6}. Targeted immunization is a highly effective vaccination strategy \cite{PS}, but it requires global information about the network thus making it impractical in real cases. Cohen $et$ $al.$ proposed acquaintance immunization strategy \cite{Imm3}, based on the immunization of a small fraction of random neighbors of randomly selected nodes. Its principle can be described as a node with higher degree is easier to be chosen from a random link. Without specific knowledge of the network, this method is efficient for networks of any broad-degree distribution, and allows for a relatively low threshold of immunization. Besides, some other novel immunization strategies have been proposed in the last decade \cite{Imm10,Imm11,Imm12,Imm13}, and have their applications in different cases.

However, many infectious diseases outbreak seasonally, which is not fully discussed in previous literature as far as we know \cite{Season,SeasonSIR}. The periodic change of temperature, humidity profiles, or even the succession of school terms and holidays, can lead to periodic phenomena of epidemics. Previous data have shown that rubella, whooping cough, and influenza reveal obvious seasonality \cite{H2000}. Here is a simple explanation of epidemics presenting seasonal phenomena: After spreading extensively, a virus dies out because infected individuals have recovered and produced antibodies. But in the next epidemic season, the mutated virus propagates again, rendering a new outbreak of the epidemic. This process then occurs repeatedly.

Herein, we propose an immunization strategy for seasonal epidemics to give a better description of this phenomenon. We merely adopt uniform immunization on the network at first. Before the start of the next epidemic season, we adjust the vaccinated nodes according to the infection status of their neighbors in previous epidemic season. This process does not need global information of the network, and achieves better performance than uniform immunization and acquaintance immunization under the same circumstances.

In this paper, we adopt susceptible-infected-recovered (SIR) model \cite{H2000,SIR1,SIR2} as epidemic spreading model, and analyze the epidemic prevalence by a heterogeneous mean-field method \cite{MF}. We introduce some parameters to investigate the evolution of vaccinated nodes during iterations, which is the main focus of the paper, and find that some nodes in the network will be selected for multiple times, although not always continuously. As epidemic season continues, the selection of vaccinated individuals tends to be stable. Those nodes include both global hubs, who possess most connections, and local hubs, who are influential in their communities. In addition, we also present some numerical results of our strategy on certain real social networks. The proposed strategy can be applied efficiently in real cases, and points out a new direction in immunization of seasonal epidemics.

The strategy will be introduced in detail in Sec.II. The analytical results and numerical simulations are presented in Sec.III and Sec. IV, respectively. Finally, conclusion is drawn in Sec.V and relevant prospect is discussed.

\section{\label{sec2} Dynamical Immunization Strategy}

We consider a connected and undirected network with $N$ nodes. In every epidemic season, there are two stages: vaccinating process and epidemic spreading process. During vaccinating process, we select some nodes by a certain strategy and vaccinate them. Vaccinated nodes can neither be infected nor spread epidemic to others. After vaccinating, the epidemic begins to spread, following the dynamics of SIR model. Each unvaccinated node in the network can be in one of the three states (S, I or R). S stands for the susceptible stage, where the individual is still healthy. They catch the disease via direct contacts with infected nodes, I, at a rate $\beta$. Infected nodes will recover and become R (recovered) nodes with rate $\mu$. Without loss of generality, we set $\mu=1$ henceforth. Initially, we set an unvaccinated node as the epidemic seed, and other unvaccinated nodes are susceptible nodes. We focus on the proportion of R nodes, $r_{\infty}$, when the epidemic process dies out in one epidemic season, i.e., when no infected nodes are left in the system.

\begin{figure}[htb]
\centerline{%
\includegraphics[width=10cm]{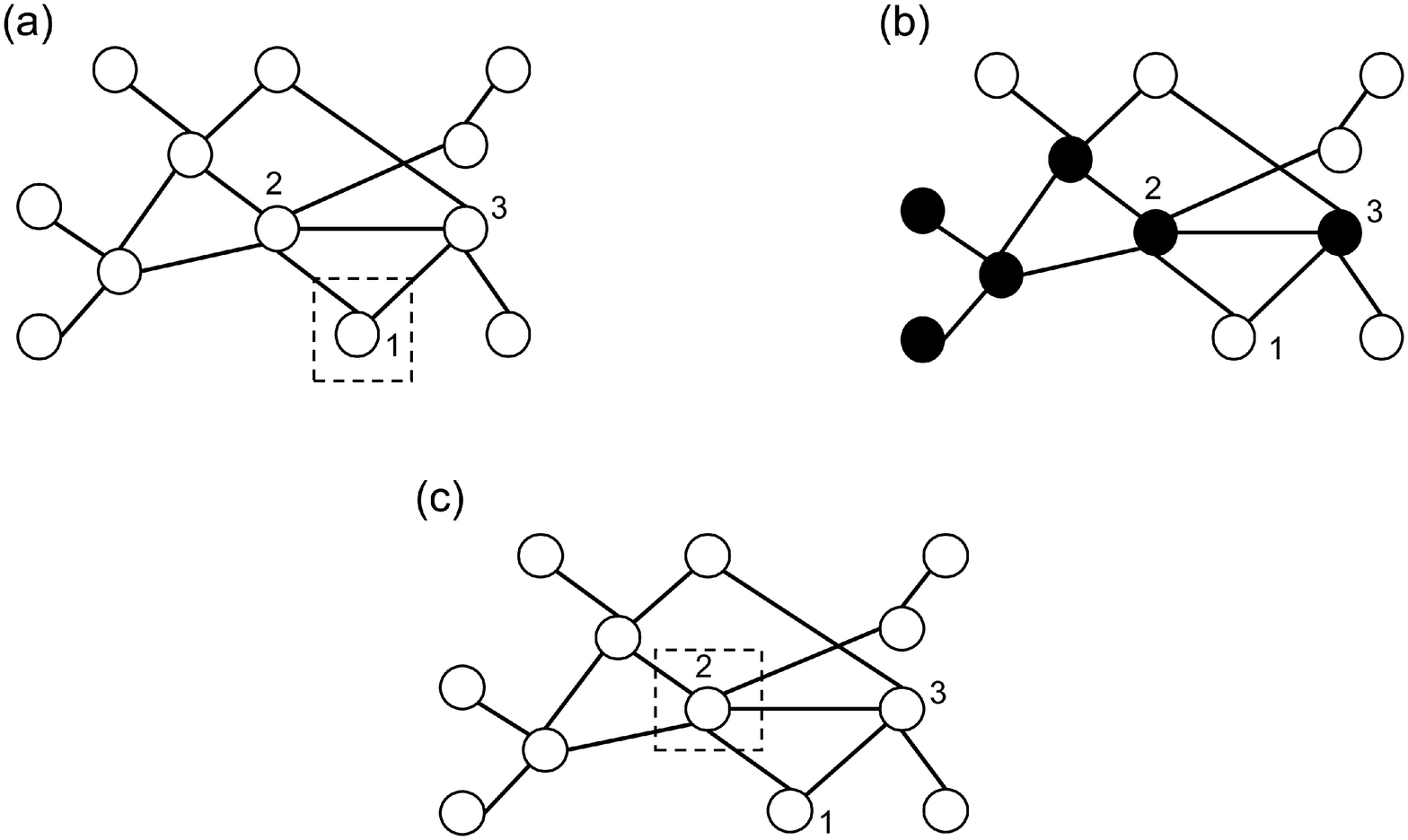}}
\caption{An illustration of our dynamical immunization strategy. First in (a), we randomly assign node 1 as the vaccinated node. After the epidemic spreading process, the system state is shown in (b). White nodes stand for susceptible nodes while black nodes for recovered nodes. We calculate the $W$ value of node 1 and its neighbors, i.e., node 2 and node 3. Since $W(1)=2$, $W(2)=3$, and $W(3)=1$, node 2 replaces node 1 as the new vaccinated node in the next epidemic season, as shown in (c).}
\label{NIll}
\end{figure}

We denote the vaccinated proportion by $v$. At the first epidemic season, $vN$ vaccinated nodes are selected randomly. That is to say, we perform uniform immunization strategy on the network initially. Let $W(u)$ stand for the infection status of node $u$ and its neighbors at the end of epidemic spreading process. We define
\begin{equation}\label{Wu}
W(u)=
\begin{cases} \sharp \{ i \text{ is R node } | i \in \Gamma (u)\} & \text{if }u \text{ is S node},
\\\sharp \{ i \text{ is R node } | i \in \Gamma (u)\}+1 &\text{if } u \text{ is R node},
\end{cases}
\end{equation}
 where $\sharp A$ denotes the number of elements in set $A$, and $\Gamma(u)$ denotes the set of $u$'s neighbors. At the second epidemic season, we adjust vaccinated nodes according to the epidemic spreading result in the first season. For every vaccinated node $u$ in epidemic season $S-1$, we calculate $W(u)$ and $W(v)$ for all $v \in \Gamma(u)$ at epidemic season $S$. Of all these nodes, we vaccinate the node who has the maximal value of $W$ since the $W$ value implies the epidemic status of the node. Thus, the new vaccinated node derived from node $u$ is $\argmax_{y \in \Gamma(u) \bigcup \{u\}}W(y)$. An illustration of our strategy is shown in Fig.\ref{NIll}.

To note, if two different vaccinated nodes ($u$ and $v$) induce the same new vaccinated node $w$ in the next epidemic season, we then select one node at random, say $u$, and set the new vaccinated node that $u$ induces as $\argmax_{y \in \Gamma(u) \bigcup \{u\}\backslash {v}}W(y)$. If no appropriate neighbor can be chosen, the vaccinated node remains unchanged.

After the selection of vaccinated nodes, the epidemic spreads again. The epidemic spreading process in this season is independent with the previous results, i.e., the epidemic seed is randomly selected from unvaccinated nodes and other unvaccinated nodes are all susceptible. After the epidemic, we then select new vaccinated nodes according to the epidemic results. In a word, vaccinating process and epidemic spreading process occur in cycles.

Obviously, we do not need the information of the entire network structure during the immunization process. Therefore, it is a local strategy.

\section{\label{sec3}Theoretical Analysis}

In this section, we establish a simple theoretical framework of our strategy. Here we adopt a heterogeneous mean-field method \cite{MF} that nodes are characterized by their degrees. The first and second moment of the network are $\langle k \rangle=\sum k P(k)$ and $\langle k^2 \rangle=\sum k^2 P(k)$, where $P(k)$ is the degree distribution of the network. The proportion of S, I, and R nodes at time $t$ is $s(t)$, $i(t)$, and $r(t)$, respectively. Due to the property of immunization, the proportion of vaccinated nodes remains unchanged. Thus, we have $s(t)+i(t)+r(t)=1-v$. Let $v_k^{(S)}$ represent the probability that a node with degree $k$ is chosen to be vaccinated at season $S$. Naturally, it holds that $\sum P(k)v_k^{(S)}=v$.

 At season $S=1$, we apply uniform immunization, i.e., $v_k^{(1)}\equiv v$. The introduction of a density $v$ of immune individuals chosen at random is equivalent to a simple rescaling of the epidemic spreading rate as $\beta$ to $\beta(1-v)$, i.e. the rate at which new infected nodes appear is decreased by a factor proportional to the probability that they are not immunized.
We denote the density of S, I, and R nodes in the degree class $k$ at epidemic season $S$ by $s_k^{(S)}$, $i_k^{(S)}$, and $r_k^{(S)}$, respectively. The SIR model evolution reads as \cite{SIREquation}
\begin{equation}\label{S1}
   \left\{
    \begin{array}{ll}
    \frac{\mathrm{d}i_k^{(1)}}{\mathrm{d}t}=\beta(1-v)ks_k^{(1)}\Theta^{(1)}-i_k^{(1)}, \\
    \frac{\mathrm{d}s_k^{(1)}}{\mathrm{d}t}=-\beta(1-v)ks_k^{(1)}\Theta^{(1)}, \\
    \frac{\mathrm{d}r_k^{(1)}}{\mathrm{d}t}=i_k^{(1)}.
    \end{array}
\right.
\end{equation}

The initial conditions are $i_k^{(1)}(0)=i_0$, $r_k^{(1)}(0)=0$, and $s_k^{(1)}(0)=1-i_0$ for any $k$. In the equation above, $\Theta^{(1)}$ stands for the average density of infected individuals at vertices pointed by any given edge at $S=1$. Assuming the network structure is uncorrelated \cite{Unc}, we have
\begin{equation}
   \Theta^{(1)}=\frac{\sum_k (k-1)P(k)i_k^{(1)}(t)}{\langle k \rangle}.
\label{Theta}
\end{equation}

By the method presented in Ref \cite{Dyna}, we can obtain that the total epidemic prevalence in season $S=1$ is \cite{SIREquation}
\begin{equation}
   r^{(1)}(\infty)=\sum_k P(k)(1-e^{-\beta(1-v)k\phi_{\infty}}),
\label{rinf}
\end{equation}
where $\phi_{\infty}$ satisfies
\begin{equation}
   \phi_{\infty}=1-\frac{1}{\langle k \rangle}-\frac{1}{\langle k \rangle}\sum_k (k-1)P(k)e^{-\beta(1-v)k\phi_{\infty}}.
\label{phiinf}
\end{equation}

Moreover, the immunization threshold (for uniform immunization) is \cite{SIREquation}
\begin{equation}
  v_c=1-\frac{\langle k \rangle}{\beta(\langle k^2 \rangle-\langle k \rangle)}.
\label{threshold}
\end{equation}

Using the numerical method, we can calculate $r_k^{(1)}$ by Eq.\ref{S1}. Since we have assumed that the network is uncorrelated, the probability of an individual has a neighbor with degree $k$ is $\eta(k)\equiv kP(k)/\sum kP(k)$. Therefore, the probability that a node with degree $k$ has $l$ recovered neighbors is $\binom{k}{l}(p^{(1)})^l (1-(p^{(1)}))^{(k-l)}$,
where $p^{(1)}=\sum_{i=1}^{k_m}\eta(i)r_i^{(1)}$ and $k_m$ is the maximal degree of the network. Due to the property of  binomial distribution, the expectation of the number of recovered neighbors that a node with degree $k$ has is $kp^{(1)}$, and the probability that the node itself is a recovered node is $r_k^{(1)}$. Thus we have $W_k^{(1)}=kp^{(1)}+r_k^{(1)}$. For convenience, we assume that the probability that a node is chosen to be the new vaccinated node is proportional to its $W$ value. Therefore, we derive that the probability of a node with degree $k$ is immunized in the epidemic season $S=2$ is
\begin{equation}
   v_k^{(2)}=\frac{(kp^{(1)}+r_k^{(1)})v}{\sum (kp^{(1)}+r_k^{(1)}) P(k)}.
\label{fk2}
\end{equation}

Then, the dynamical equation of epidemic spreading in season $S=2$ is
\begin{equation}\label{S2}
   \left\{
    \begin{array}{ll}
    \frac{\mathrm{d}i_k^{(2)}}{\mathrm{d}t}=\beta(1-v_k^{(2)})ks_k^{(2)}\Theta^{(2)}-i_k^{(2)}, \\
    \frac{\mathrm{d}s_k^{(2)}}{\mathrm{d}t}=-\beta(1-v_k^{(2)})ks_k^{(2)}\Theta^{(2)}, \\
    \frac{\mathrm{d}r_k^{(2)}}{\mathrm{d}t}=i_k^{(2)}.
    \end{array}
\right.
\end{equation}

Similarly, we can obtain $r_k^{(2)}$ for $k=1,2,...k_m$ and $r_{\infty}^{(2)}=\sum_k P(k)r_k^{(2)}$. The iterative process can be applied recursively and a series of $r_{\infty}^{(S)}$ ($S=2,3,...$) can be obtained by Eq.\ref{iter}:

\begin{equation}\label{iter}
   \left\{
    \begin{array}{ll}
    \frac{\mathrm{d}i_k^{(S)}}{\mathrm{d}t}=\beta(1-v_k^{(S)})ks_k^{(S)}\Theta^{(S)}-i_k^{(S)}, \\
    \frac{\mathrm{d}s_k^{(S)}}{\mathrm{d}t}=-\beta(1-v_k^{(S)})ks_k^{(S)}\Theta^{(S)}, \\
    \frac{\mathrm{d}r_k^{(S)}}{\mathrm{d}t}=i_k^{(S)},
    \end{array}
\right.
\end{equation}

where $v_k^{(S)}$ satisfies
\begin{equation}
   v_k^{(S)}=\frac{(kp^{(S-1)}+r_k^{(S-1)})v}{\sum (kp^{(S-1)}+r_k^{(S-1)}) P(k)},
\label{vks}
\end{equation}
 and $p^{(S-1)}=\sum_{i=1}^{k_m}\eta(i)r_i^{(S-1)}$.

\begin{figure}[htb]
\centerline{%
\includegraphics[width=10cm]{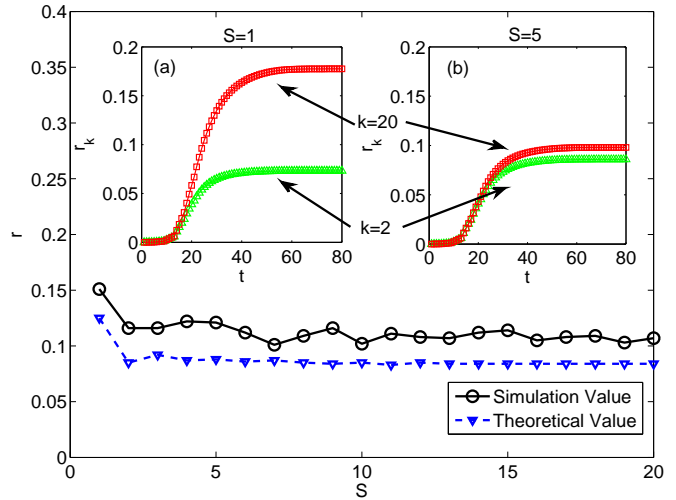}}
\caption{Main plot: The comparison between simulation values and theoretical values for $r_{\infty}^{(S)}$. The black circles denote simulation value while the blue dashed line for theoretical value. Inset: The relationship of $r_k^{(S)}$ and time step $t$. Left figure is for $S=1$ while right figure is for $S=5$. Top curves (red squares) stand for $k=20$, and bottom curves (green triangles) stand for $k=2$. Here we adopt BA network with $N=100$ and $\langle k \rangle=3.94$. Epidemic parameters are $\beta=0.1$ and $v=0.1$. Each simulation point is the average value of $10^2$ experiments.}
\label{NSiTh}
\end{figure}

Due to the complexity of the numerical algorithm, here we adopt a BA network \cite{BA} with $N=100$ to verify our results. BA network is a classic heterogeneous network topology, and has highly skewed degree distribution. In the main plot of Fig.\ref{NSiTh}, we compare the simulation value with the theoretical value. The numerical results are relatively consistent with the simulation. The inset figures present the relationship of $r_k^{(S)}(t)$ and time step $t$. Left figure is for $S=1$ while right figure is for $S=5$. Top curves (red squares) stand for $k=20$, and bottom curves (green triangles) stand for $k=2$. Compared with uniform immunization ($S=1$), our method significantly decreases the epidemic prevalence for higher degree nodes.

\section{\label{sec4}Simulation Results}

In this section, we further investigate our strategy by numerical simulations. Here we adopt four real networks \cite{Stanford}. (1) Wikipedia Vote Network: Wikipedia is a free encyclopedia written collaboratively by volunteers around the world. Nodes in the network represent wikipedia users and a directed edge from node $i$ to node $j$ represents that user $i$ voted on user $j$. (2) Epinions Social Network: This is a who-trust-whom online social network of a general consumer review site Epinions.com. Nodes in the network represent Epinions users and edges represent trust relationship. (3) Slashdot Social Network: Slashdot is a technology-related news website known for its specific user community. The network contains links between the users of Slashdot. (4) Enron Email Network: Enron email communication network covers email communication within a dataset emails. Nodes of the network are email addresses and if an address $i$ sent at least one email to address $j$, the graph contains an edge from $i$ to $j$.

\begin{table}
\centering
\caption{The statistics of four real networks.\label{table1}}
\begin{tabular}{lcccccc}
\hline
\hline
  Name &$N$ &$\langle C \rangle$ &$\langle k \rangle$ & $\langle k^2 \rangle$ & $v_c$ ($\beta=0.1)$ & $v_c$ ($\beta=0.05$) \\\hline

   Wiki-Vote  & 7115 &0.14 & 29.4 & 4554.8 & 0.935 & 0.870\\
  Epinions & 75879  &0.14 & 16.4 & 3172.1 & 0.955 & 0.909 \\
  Slashdot & 77360  &0.06 & 23.5 & 6428.8 & 0.963 & 0.927 \\
  Enron & 36692  &0.50 & 22.5 & 6812.1 & 0.967 & 0.934 \\

  \hline
  \hline
\end{tabular}
\end{table}

\begin{figure}[htb]
\centerline{%
\includegraphics[width=10cm]{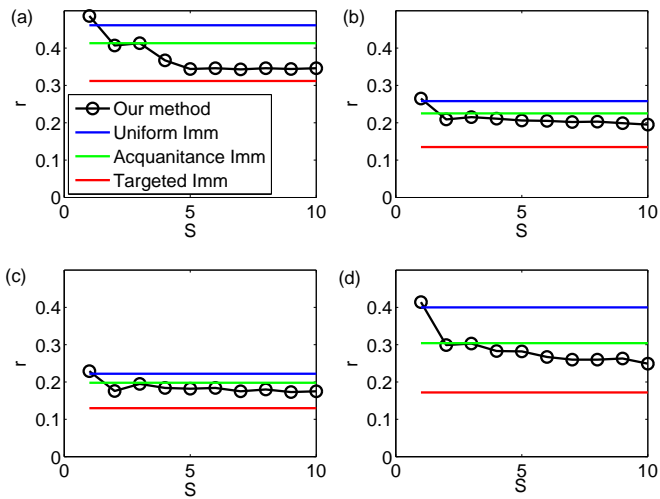}}
\caption{The relationship of epidemic prevalence $r_{\infty}^{(S)}$ and epidemic season $S$ for different immunization strategies. Black circles stand for our dynamical immunization method. Three solid lines represent uniform immunization (blue), acquaintance immunization (green), and targeted immunization (red) from top to bottom. Results of Wiki-Vote network, Epinions network, Slashdot network, and Enron network are shown in (a), (b), (c), and (d), respectively. Epidemic parameters are $\beta=0.1$ and $v=0.1$. Each point is the average value of $10^2$ experiments.}
\label{Nrs}
\end{figure}

Although these four networks are online social networks or email networks, they have similar spreading properties with actual inter-personal networks that are hard to be obtained \cite{PS46,Muchnik2013,LN2008}. The statistics of these four networks are listed in Table.\ref{table1}, where $\langle C \rangle$ stands for the average clustering coefficient of the network \cite{Newman2003}. We treat all directed links as undirected links for convenience. It is shown in the table that $v_c$ is too large to be obtained, implying that the total immunity is hard to reach. It is a reflection of the finding that heterogeneous network is an ideal substrate for epidemic spreading \cite{Spreading,Boguna2003}.

\begin{figure}[htb]
\centerline{%
\includegraphics[width=8cm]{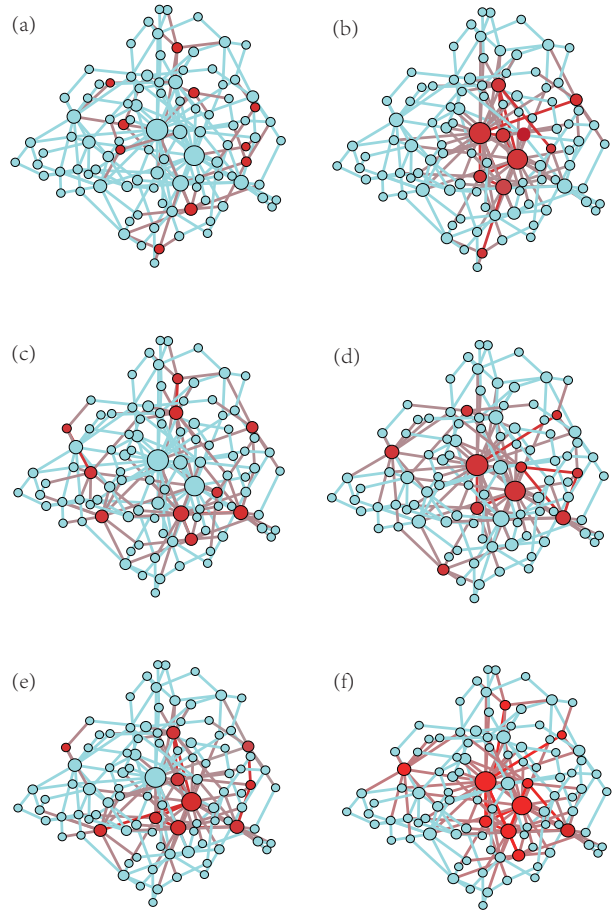}}
\caption{A demonstration of immunization for $S=1$ to $6$ ((a) to (f)). Here we adopt a BA network with $N=100$ and $\langle k \rangle=3.94$. Epidemic parameters are $\beta=0.1$ and $v=0.1$. The size of the node implies its degree. Red (dark) nodes stand for vaccinated nodes while blue (light) nodes stand for unvaccinated nodes. We merely show the immunization situation, neglecting the epidemic spreading results.}
\label{NSixFig}
\end{figure}

Naturally, we compare our method with other classic immunization strategies such as uniform immunization, targeted immunization, and acquaintance immunization. Note that these immunization strategies are independent of the epidemic season. As is shown in Fig.\ref{Nrs}, the epidemic prevalence of our method gradually decreases to a steady state. Generally, our method has a better performance than uniform immunization and acquaintance immunization after a few seasons. Though targeted immunization performs the best, the fact that it needs global information makes it impractical in real cases. Our strategy compromises between immunization efficiency and limited information. As it can be seen from Fig.\ref{Nrs}, the curves of our method do not decrease monotonically from the first season. The fluctuation can be interpreted as follows. In a particular epidemic season $S$, some influential spreaders (i.e., who play important roles in the spreading process) are vaccinated. Therefore it is relatively difficult for their neighbors to be infected in the next epidemic season $S+1$, and they are not chosen to be the vaccinated nodes due to the fact that they have fewer infected neighbors. While the new vaccinated nodes are less influential than those nodes, so the infected populations of epidemic season $S+1$ increase. In the epidemic season $S+2$, those influential nodes are set to be vaccinated nodes again, which lessens the risk of the epidemic.

\begin{figure}[htb]
\centerline{%
\includegraphics[width=10cm]{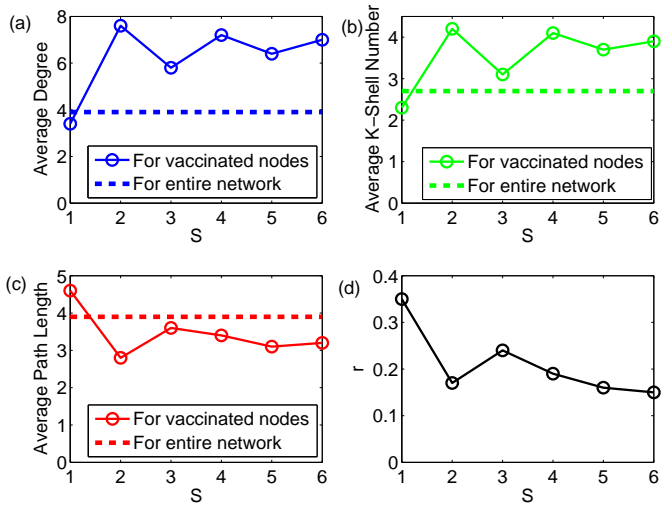}}
\caption{The curves of average degree (a), average k-shell (b), and average path length(c) in the BA network. The dashed lines represent the value of the entire network, and the circles represent the corresponding value among vaccinated nodes. The epidemic prevalence is shown in (d). Network structure is shown in Fig.\ref{NSixFig}. Epidemic parameters are $\beta=0.1$ and $v=0.1$.}
\label{BAPara}
\end{figure}

\begin{figure}[htb]
\centerline{%
\includegraphics[width=10cm]{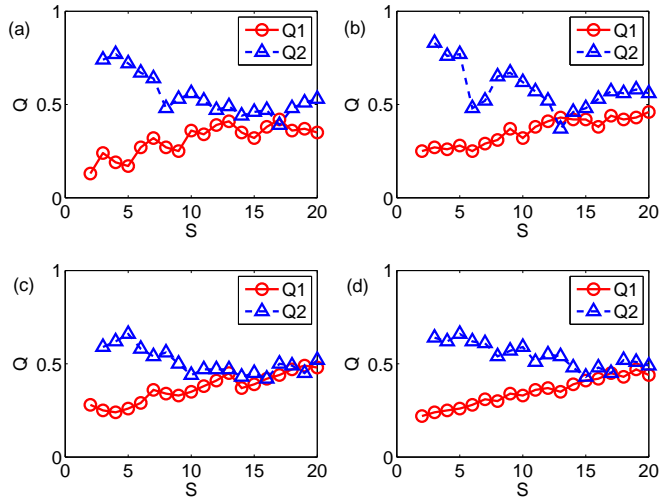}}
\caption{The relationship of recurrence rate $Q_1(S)$ ($Q_2(S)$) and epidemic season $S$. The red circles stand for $Q_1(S)$ and the blue triangles for $Q_2(S)$. Results of Wiki-Vote network, Epinions network, Slashdot network, and Enron network are shown in (a), (b), (c), and (d), respectively. Epidemic parameters are $\beta=0.1$ and $v=0.1$. Each point is the average value of $10^2$ experiments.}
\label{CFL}
\end{figure}

However, the selection of vaccinated nodes is not a periodic process, but rather an optimizing process. Fig.\ref{NSixFig} shows the evolution of vaccinated nodes in a BA network. The red (dark) nodes denote vaccinated nodes while the blue (light) nodes denote unvaccinated nodes. Fig.\ref{BAPara} presents some statistical properties of the entire network and the vaccinated nodes from $S=1$ to $S=6$, including the averaged degree, the average k-shell number \cite{kshell} and the average path length \cite{Dyna}. The degree and the k-shell number describe the importance of a node, and the average path length among nodes implies their closeness \cite{PS}. As illustrated in Fig.\ref{BAPara}, for vaccinated nodes, the average degree and the average k-shell obtain their maximal values (average path length obtain its minimal value) at $S=2$, but the epidemic prevalence obtains its minimal value at $S=6$. It shows that to prevent the epidemic, we do not need to vaccinate the nodes who have the highest degree values or k-shell numbers. In other words, if we rank the nodes according to their degrees, the best immunization strategy is not vaccinating nodes in this sequence. Comparing Fig.\ref{NSixFig}(b) with Fig.\ref{NSixFig}(d), we may find that hubs are prone to be selected as vaccinated nodes at the second epidemic season. Nevertheless, it is not the best status because vaccinated nodes are too centralized, and numerous peripheral nodes are not covered. The situation is improved as the evolution continues. As can be seen from Fig.\ref{NSixFig} (f), at the sixth epidemic season, some ``local hubs'', which do not have the highest degree value but may connect some communities or peripheral nodes closely, are included. Thus, our method not only concerns connections or k-shell numbers of a node, but also pays attention to the nodes who have important locations in the network \cite{Kitsak2010,PS2014}.

Next, we introduce recurrence rate $Q_1(S)$ and $Q_2(S)$ to investigate the overlapping of vaccinated nodes. Here $Q_1(S)$ is the proportion of nodes vaccinated in both epidemic seasons $S$ and $S-1$ ($S=2,3,...$), and $Q_2(S)$ is the proportion of nodes vaccinated in both epidemic seasons $S$ and $S-2$ ($S=3,4,...$). For example, if there are $G$ common nodes that are vaccinated both in epidemic seasons $S$ and $S-1$, then $Q_1(S)=G/(vN)$. These parameters reflect the fluctuation of the evolution and are graphically presented in Fig.\ref{CFL}. As epidemic season continues, $Q_1(S)$ rises while $Q_2(S)$ shows a decreasing trend in general. But when $S$ is large enough, $Q_2(S)$ is prone to increase and finally maintains a relatively steady level. It implies that some nodes have a higher probability of being vaccinated repeatedly.

\begin{figure}[htb]
\centerline{%
\includegraphics[width=10cm]{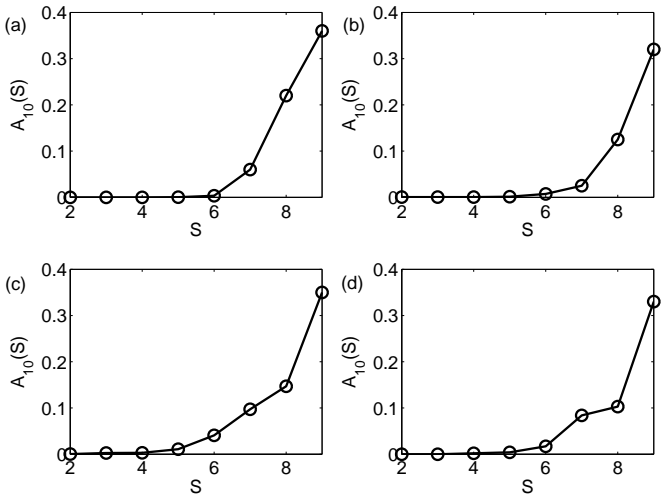}}
\caption{The relationship of $A_{10}(S)$ and epidemic season $S$. Results of Wiki-Vote network, Epinions network, Slashdot network, and Enron network are shown in (a), (b), (c), and (d), respectively. Epidemic parameters are $\beta=0.1$ and $v=0.1$. Each point is the average value of $10^2$ experiments.}
\label{BHL}
\end{figure}

Fig.\ref{CFL} shows the overlapping of vaccinated nodes in epidemic seasons $S$ and $S-1$ ($S-2$), but does not provide information about how many nodes are vaccinated continuously. To illustrate this issue clearly, we introduce $A_{S'}(S)$, which is defined as the proportion of nodes that have been immunized continuously from epidemic season $S$ to $S'$ ($S<S'$). Plotting $A_{10}(S)$ for $S=2,3,...9$ yields the curve in Fig.\ref{BHL}. Here we set $S'=10$, which is large enough for real cases (the interval between two epidemics may be a year or so in reality). We neglect the epidemic season $S=1$ because it is a uniform immunization. As it can be seen from Fig.\ref{BHL}, there are few nodes that have been vaccinated for more than four continuous epidemic seasons.

\begin{figure}[htb]
\centerline{%
\includegraphics[width=10cm]{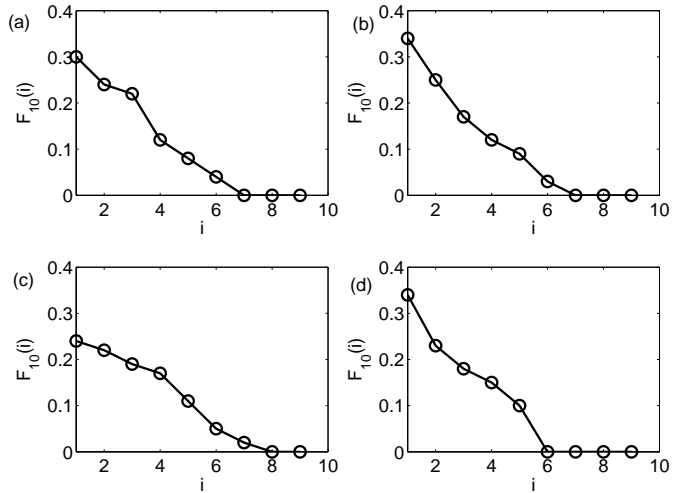}}
\caption{The relationship of $F_{10}(i)$ and the number of epidemic seasons $i$. Results of Wiki-Vote network, Epinions network, Slashdot network, and Enron network are shown in (a), (b), (c), and (d), respectively. Epidemic parameters are $\beta=0.1$ and $v=0.1$. Each point is the average value of $10^2$ experiments.}
\label{FS}
\end{figure}

Although the number of nodes that have been vaccinated continuously is extremely small, there are some individuals that have been vaccinated many times (shown in Fig.\ref{FS}). Here $F_{S'}(i)$ is defined as the ratio of the number of nodes that have been immunized for $i$ times (not necessarily continuously) from the second epidemic season to epidemic season $S'$ ($S'>2$) to the total number of nodes that have been immunized for at least one epidemic season. The first epidemic season is also neglected because of its randomness. Here we also set $S'=10$. Naturally, we have $F_{10}(1)+F_{10}(2)+...+F_{10}(9)=1$. In Fig.\ref{FS}, we can find that more than $10\%$ nodes are vaccinated for four epidemic seasons. Combining Fig.\ref{BHL} and Fig.\ref{FS}, we may draw the conclusion that there are some ``core individuals'' in the evolution.  In every epidemic season, the vaccinated nodes are a group of ``core individuals'' (some global hubs and some local hubs), with the existence of some other nodes. Different ``core groups'' emerge repeatedly but not continuously.

\begin{figure}[htb]
\centerline{%
\includegraphics[width=10cm]{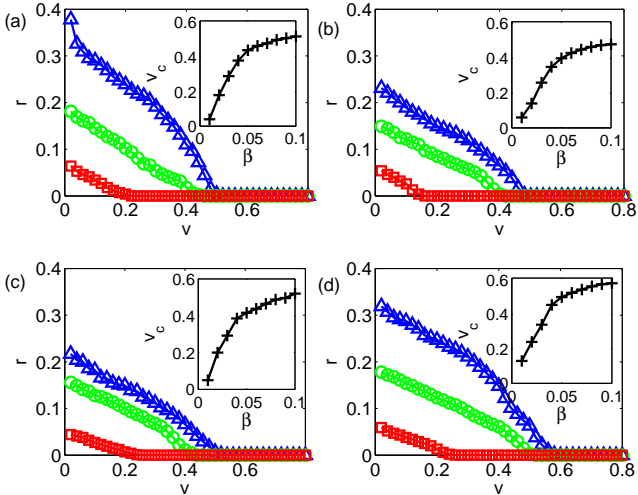}}
\caption{Main plot: The relationship of epidemic prevalence $r_{\infty}^{(S)}$ and immunization proportion $v$ for different infected rate $\beta$. Here $S=5$. Blue triangles, green circles, and red squares stand for the case of $\beta=0.10$, $\beta=0.05$, and $\beta=0.02$, respectively. Inset: The relationship of vaccinated threshold $v_c$ and $\beta$. Here $v_c$ is the value of immunization opinion that makes $r_{\infty}^{(5)}<0.005$. Results of Wiki-Vote network, Epinions network, Slashdot network, and Enron network are shown in (a), (b), (c), and (d), respectively. Each point is the average value of $10^2$ experiments.}
\label{NNrv}
\end{figure}

Now, we concentrate on the critical vaccinated proportion that can halt virus. The $v_c$ obtained in Eq.\ref{threshold} is actually for uniform immunization and is extremely large (as shown in Table.\ref{table1}). Here we fix infected rate $\beta$ and vary immunization proportion $v$ to observe the epidemic prevalence $r_{\infty}$. As the main plot of Fig.\ref{NNrv} suggests, the curves stand for the relationship between $r$ and $v$ is almost linear at initial and then drops to zero. Obviously, increasing the immunization proportion is an effective method to diminish the epidemic prevalence. In the inset plot of Fig.\ref{NNrv}, we draw the curve of vaccinated threshold $v_c$ that varies by the infected rate $\beta$. It provides an estimation of immunization proportion that inhibits epidemics totally. Obviously, the vaccinated threshold of our method decreases dramatically compared with the uniform cases.

\section{\label{sec5}Conclusion}

To summarize, we have proposed an model to describe seasonal epidemics and have presented related immunization strategy. The selection of vaccinated nodes is optimized gradually, based on local information of the network from the previous epidemic season. We establish the theoretical framework of our model, which is in agreement with the simulation. We also compare our method with other immunization strategies and find that our method performs superiorly to uniform immunization and acquaintance immunization. These findings suggest that our method provides useful hints for immune strategy design. Meanwhile, we discuss the evolution of vaccinated individuals, and find that the influential nodes are vaccinated repeatedly but not continuously. As epidemic season continues, the selection of vaccinated nodes tends to remain stable, and locates on both global hubs and local hubs. We also present the relationship between epidemic prevalence and immunization proportion numerically.

In further research, we would like to discuss the relationship of epidemic prevalence and the network structure by our strategy, especially for the community structure and correlated networks \cite{Newman2003}. Besides, the property of the ``core individuals'' is also an interesting topic that can be studied in more depth.

\begin{acknowledgments}
This work is supported by Major Program of National Natural Science Foundation
of China (11290141), NSFC (11201018), International Cooperation Project no.
2010DFR00700, Fundamental Research of Civil Aircraft no. MJ-F-2012-04, and Beihang University Innovation And Practice Fund For Graduates.
\end{acknowledgments}

\nocite{*}

\bibliography{apssamp}

\end{document}